\documentclass[review,12pt]{elsarticle}

\usepackage[misc]{ifsym}
\usepackage[colorinlistoftodos]{todonotes}

\usepackage[caption=false]{subfig}
\usepackage{float}

\usepackage{multirow}
\usepackage{rotating}

\usepackage{enumitem}

\usepackage{lineno,hyperref}

\usepackage[small]{caption}

\usepackage{siunitx}
\usepackage{pifont}
\newcommand{\cmark}{\ding{51}}%
\newcommand{\xmark}{\ding{55}}%

\usepackage{mathtools,etoolbox}
\DeclarePairedDelimiterX{\abs}[1]{\lvert}{\rvert}{\ifblank{#1}{{}\cdot{}}{#1}}

\usepackage{graphicx}
\usepackage{amssymb}
\usepackage{amsmath}

\usepackage{lineno}
\usepackage{soul}
\usepackage{manyfoot}

\DeclareNewFootnote{A}
\DeclareNewFootnote{B}

\makeatletter
\def\Hline{
\noalign{\ifnum0=`}\fi\hrule \@height 1pt \futurelet
\reserved@a\@xhline}
\makeatother

\biboptions{sort&compress}

\journal{Computers in Biology and Medicine}

\begin{document}


\begin{titlepage}
	\clearpage\thispagestyle{empty}
		
\begin{center}
\begin{Large}Focus U-Net: A novel dual attention-gated CNN for polyp segmentation during colonoscopy\end{Large}
\end{center}

\noindent Michael Yeung\footnote[1]{Corresponding author. \textit{Phone}: +44 (0) 1223 256255.\\  \textit{Address}: Department of Radiology, University of Cambridge, Box 218, Cambridge Biomedical Campus, Cambridge CB2 0QQ, United Kingdom.}, \textit{E-mail}: \texttt{mjyy2@cam.ac.uk}\\
\begin{scriptsize}
Department of Radiology, University of Cambridge, Cambridge CB2 0QQ, United Kingdom\\
School of Clinical Medicine, University of Cambridge, Cambridge CB2 0SP, United Kingdom
\end{scriptsize}

\noindent Evis Sala, \textit{E-mail}: \texttt{es220@medschl.cam.ac.uk}\\
\begin{scriptsize}
Department of Radiology, University of Cambridge, Cambridge CB2 0QQ, United Kingdom\\
Cancer Research UK Cambridge Centre, University of Cambridge, Cambridge CB2 0RE, United Kingdom
\end{scriptsize}

\noindent Carola-Bibiane Sch\"{o}nlieb, \textit{E-mail}: \texttt{cbs31@cam.ac.uk}\\
\begin{scriptsize}
Department of Applied Mathematics and Theoretical Physics, University of Cambridge, Cambridge CB3 0WA, United Kingdom
\end{scriptsize}

\noindent Leonardo Rundo, \textit{E-mail}: \texttt{lr495@cam.ac.uk}\\
\begin{scriptsize}
Department of Radiology, University of Cambridge, Cambridge CB2 0QQ, United Kingdom\\
Cancer Research UK Cambridge Centre, University of Cambridge, Cambridge CB2 0RE, United Kingdom
\end{scriptsize}

\newpage
\end{titlepage}


\begin{frontmatter}


\title{Focus U-Net: A novel dual attention-gated CNN for polyp segmentation during colonoscopy}




\author[CamRadiol,CamMedSchool]{Michael Yeung\corref{corr}}

\author[CamRadiol,CRUKCC]{Evis Sala}

\author[DAMTP]{\\Carola-Bibiane Sch\"{o}nlieb}

\author[CamRadiol,CRUKCC]{Leonardo Rundo}

\cortext[corr]{Corresponding author. \textit{Phone}: +44 (0) 1223 256255. \textit{Address}: Department of Radiology, University of Cambridge, Box 218, Cambridge Biomedical Campus, Cambridge CB2 0QQ, United Kingdom. \textit{E-mail}: \texttt{lr495@cam.ac.uk}}

\address[CamRadiol]{Department of Radiology, University of Cambridge,\\Cambridge CB2 0QQ, United Kingdom}
\address[CamMedSchool]{School of Clinical Medicine, University of Cambridge,\\Cambridge CB2 0SP, United Kingdom}
\address[CRUKCC]{Cancer Research UK Cambridge Centre, University of Cambridge,\\Cambridge CB2 0RE, United Kingdom}
\address[DAMTP]{Department of Applied Mathematics and Theoretical Physics, University of Cambridge, Cambridge CB3 0WA, United Kingdom}

\begin{abstract}

\noindent\textit{Background:} Colonoscopy remains the gold-standard screening for colorectal cancer. However, significant miss rates for polyps have been reported, particularly when there are multiple small adenomas. This presents an opportunity to leverage computer-aided systems to support clinicians and reduce the number of polyps missed.

\noindent\textit{Method:} In this work we introduce the Focus U-Net, a novel dual attention-gated deep neural network, which combines efficient spatial and channel-based attention into a single Focus Gate module to encourage selective learning of polyp features. The Focus U-Net incorporates several further architectural modifications, including the addition of short-range skip connections and deep supervision. Furthermore, we introduce the Hybrid Focal loss, a new compound loss function based on the Focal loss and Focal Tversky loss, designed to handle class-imbalanced image segmentation. For our experiments, we selected five public datasets containing images of polyps obtained during optical colonoscopy: CVC-ClinicDB, Kvasir-SEG, CVC-ColonDB, ETIS-Larib PolypDB and EndoScene test set. We first perform a series of ablation studies and then evaluate the Focus U-Net on the CVC-ClinicDB and Kvasir-SEG datasets separately, and on a combined dataset of all five public datasets. To evaluate model performance, we use the Dice similarity coefficient (DSC) and Intersection over Union (IoU) metrics.

\noindent\textit{Results:}  Our model achieves state-of-the-art results for both CVC-ClinicDB and Kvasir-SEG, with a mean DSC of $0.941$ and $0.910$, respectively. When evaluated on a combination of five public polyp datasets, our model similarly achieves state-of-the-art results with a mean DSC of $0.878$ and mean IoU of $0.809$, a $14\%$ and $15\%$ improvement over the previous state-of-the-art results of $0.768$ and $0.702$, respectively. 

\noindent\textit{Conclusions:} This study shows the potential for deep learning to provide fast and accurate polyp segmentation results for use during colonoscopy. The Focus U-Net may be adapted for future use in newer non-invasive colorectal cancer screening and more broadly to other biomedical image segmentation tasks similarly involving class imbalance and requiring efficiency.

\end{abstract}



\end{frontmatter}

\section{Introduction}
\label{sec:Intro}

Globally, colorectal cancer (CRC) ranks third in terms of incidence, and second only to lung cancer as a leading cause of cancer death \cite{bray2018global}. The absence of specific symptoms in the early stages of disease often results in delays in diagnosis and treatment, with the stage of disease at diagnosis strongly linked to prognosis. In the United States, the 5-year relative survival rate for Stage I colon cancer is $92\%$, decreasing to $12\%$ in those with Stage IV \cite{rawla2019epidemiology}.

In 1988, Vogelstein proposed the adenoma-carcinoma sequence model for CRC carcinogenesis, describing the transition from benign adenoma to adenocarcinoma with associated well-defined histology at each stage \cite{vogelstein1988genetic}. Importantly, there is a prolonged, identifiable and treatable preclinical phase lasting years prior to malignant transformation \cite{kuntz2011systematic, brenner2007risk}. As a result, CRC is highly suitable for population level screening, which has been shown to be effective at reducing overall mortality \cite{schreuders2015colorectal, arnold2017global}. 

Non-invasive CRC screening tests include stool-based tests, such as the faecal occult blood test, and more recent blood-based tests, such as Epi proColon\textsuperscript{®} (Epigenomics AG, Berlin, Germany). Capsule colon endoscopy and CT colonography are newer, non-invasive radiological investigations useful for screening high-risk individuals unsuitable for colonoscopy. Invasive options include flexible sigmoidoscopy and colonoscopy, offering direct visualisation and the ability to obtain biopsy specimens for histological analysis. Sigmoidoscopy is limited to cancer in the rectum, sigmoid and descending colon, and colonoscopy remains the gold-standard screening tool for CRC with the highest sensitivity and specificity \cite{issa2017colorectal}. However, colonoscopy is associated with significant miss rates for polyp detection, contributed by both patient and polyp-related factors \cite{kim2017miss, van2006polyp, leufkens2012factors}. The risk of missing polyps significantly increases in patients with two or more polyps, with higher miss rates for flat or sessile compared to pedunculated or sub-pedunculated polyps and miss rates vary from $2\%$ for adenomas $\geq 10$mm to $26\%$ for adenomas $< 5$mm.	

The difficulty in detecting polyps during colonoscopy presents an opportunity to incorporate computer-aided systems to reduce polyp miss rates \cite{le2020application}. Polyps may remain hidden from the field of view, for which a real-time Artificial Intelligence (AI) model has been developed to assess the quality of colonoscopy \cite{thakkar2020use}. Alternatively, polyps may enter the field of view but remain undetected by the operator. In this case, polyp segmentation approaches not only aim to detect polyps, but to also accurately delineate the polyp border from surrounding mucosa. Early automated methods to segment polyps relied on hand-crafted feature extraction, using either shape-based \cite{krishnan1998intestinal, hwang2007polyp, bernal2012towards} or texture and colour-based analysis \cite{karkanis2003computer, coimbra2006mpeg}. While considerable advancements were made, the accuracy of polyp segmentation remained low with hand-crafted features unable to capture the scale of polyp heterogeneity \cite{yu2016integrating}.

This paper is structured as follows.
Section~\ref{sec:Background} outlines the state-of-the-art of polyp segmentation on colonoscopy images.
Section~\ref{sec:FocusUnet} describes the architecture of the proposed Focus U-Net.
Section~\ref{sec:MaterialsAndEvalMethods} describes the analysed datasets and the evaluation metrics used in this study.
Section~\ref{sec:Results} present the experimental results.
Finally, Section~\ref{sec:Discussion} provides a discussion and concluding remarks.

\section{Related work}
\label{sec:Background}

In recent years, significant improvements have been achieved by adopting automatic methods based on deep learning. The introduction of Fully Convolutional Networks (FCN) enabled Convolutional Neural Network (CNN) architectures to tackle semantic image segmentation tasks \cite{long2015fully}. The application of FCNs to polyp segmentation has yielded impressive results \cite{akbari2018polyp, brandao2017fully}. Currently, the state-of-the-art approaches are largely based on the U-Net, a modified FCN architecture developed for biomedical image segmentation \cite{ronneberger2015u}. The U-Net consists of an encoding network used to capture the image context, followed by a symmetrical decoding network enabling localisation of salient regions. UNet++ extends the U-Net by incorporating a series of nested skip connections, reducing the semantic gap between the features maps of the encoder and decoder networks prior to fusion \cite{zhou2019unet++, jha2019resunet++, jha2021comprehensive}. The ResUNet++ combines residual units with the spatial attention-based Atrous Spatial Pyramidal Pooling (ASPP) and channel attention-based squeeze-and-excitation block \cite{jha2019resunet++, jha2021comprehensive}. Similarly, both attention components are incorporated into the DoubleU-Net, which further leverages transfer learning from the first U-Net to generate features as input into the second network \cite{jha2020doubleu}. Despite excellent segmentation results with these models, the large memory and associated long inference time limits use in clinical practice where real-time polyp segmentation is required. Recently, several efficient models with significantly faster inference times, in addition to greater accuracy, have been proposed. PolypSegNet introduces the depth dilated inception module, enabling efficient feature extraction across a range of receptive field sizes \cite{mahmud2021polypsegnet}. PraNet uses a two-step process that involves initial localisation of the polyp area, followed by progressive refining of the polyp boundary, resembling the method by which humans identify polyps \cite{fan2020pranet}. HarDNet-MSEG uses a low memory latency HarDNet68 backbone \cite{huang2021hardnet}, together with a Cascaded Partial Decoder \cite{chao2019hardnet} for fast and accurate polyp segmentation. 

In this paper, we introduce a novel attention-gated U-Net architecture, named the Focus U-Net, which uses a new attention module known as the Focus Gate (FG), incorporating both spatial and channel-based attention with a focal parameter to control the degree of background suppression. Using this architecture, we achieve state-of-the-art results across five public polyp segmentation datasets. With an efficient and accurate polyp segmentation algorithm, we provide the latest advancement towards using AI in colonoscopy practice, with the aim of assisting clinicians by increasing polyp detection rates.

\section{The proposed Focus U-Net architecture}
\label{sec:FocusUnet}

In this section, we introduce the techniques used in the Focus U-Net, beginning with the FG and associated channel and spatial attention modules, followed by explanations of deep supervision and loss function optimisation. 

\subsection{Overview of the Focus U-Net}

\begin{figure}[ht!]
    \centering
    \hspace*{-2.6cm}
    \includegraphics[width=1.4\textwidth]{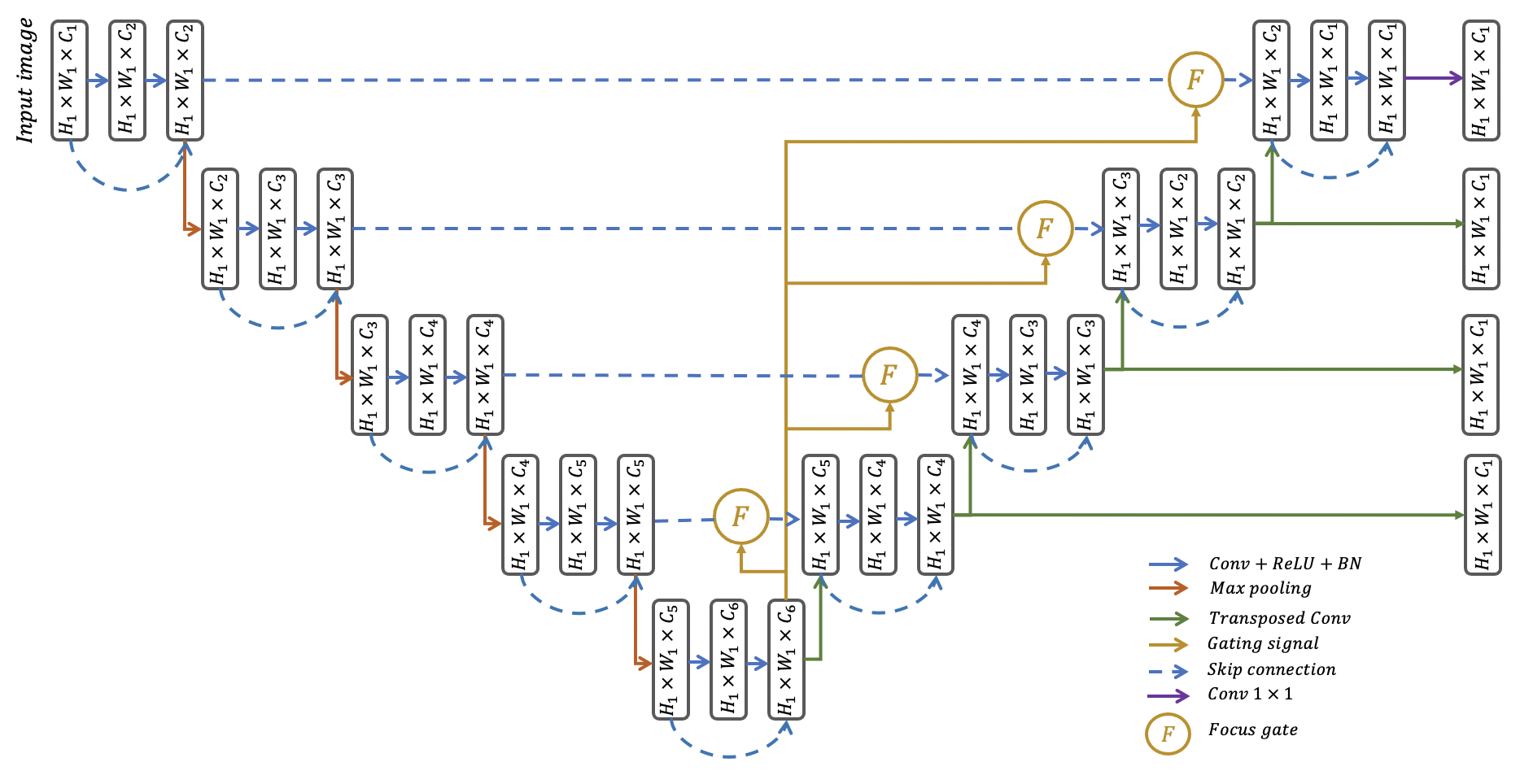}
    \caption{Architecture of the proposed Focus U-Net. The gating signal originates from the deepest layer in the network and refines incoming skip connection input at each depth. Deep supervision is represented by the dark green arrows generating outputs at each depth. }
    \label{fig:figure_1}
\end{figure}

The architecture of the Focus U-Net is shown in Fig.~\ref{fig:figure_1}. Similar to the U-Net, the Focus U-Net begins with an encoding network, capturing features relevant to polyps such as edges, texture and colour. The deepest layer of the network contains the richest information relating to image features, at the cost of spatial resolution, and forms the gating signal used as input into the FG. The FG uses the gating signal to refine incoming signals from the encoding network in the form of long-range skip connections, by highlighting specific image features and regions that are integrated into the decoding network. Successive upsampling in the decoding network enables polyp localisation at progressively higher resolution, with the final output producing the segmentation map defining, if present, the precise shape and location of the polyp. Short-range skip connections and deep supervision create additional pathways for information transfer, diversifying the features extracted and providing shortcuts for the loss to propagate backwards to the deeper layers when updating parameters.

\subsection{Attention Gates and the Focus Gate}

The concept of attention mechanisms in neural networks is inspired by cognitive attention, where relevant stimuli in the visual field are identified and selectively processed. In the context of neural networks, distinctions are made between hard and soft attention, as well as global and local attention \cite{han2021madgan,chen2021iba}.
Hard attention calculates attention scores for each region of the image to select the regions to attend. This requires a stochastic sampling process, which is a non-differentiable calculation relying on reinforcement learning to update parameters \cite{tagliarini1987neural}. In contrast, soft attention is deterministic and assigns regions of interest (ROIs) with higher weight, with the benefit that this process is differentiable and therefore trainable by standard backpropagation \cite{jetley2018learn, wang2017residual}. The distinction between global and local attention refers to whether the whole input or only a subset of the input is attended \cite{luong2015effective}. For training of neural networks, a combination of soft and local attention is often favoured \cite{schlemper2019attention}. 

Attention Gates (AGs) provide neural networks with the capacity to selectively attend to inputs. The use of AG first originated in the context of machine translation as part of Natural Language Processing (NLP) \cite{luong2015effective}, \cite{bahdanau2014neural, shen2018disan, vaswani2017attention}, but has also more recently shown success in Computer Vision \cite{schlemper2019attention}. 

\begin{figure}[ht!]
    \centering
    \includegraphics[width=0.8\textwidth]{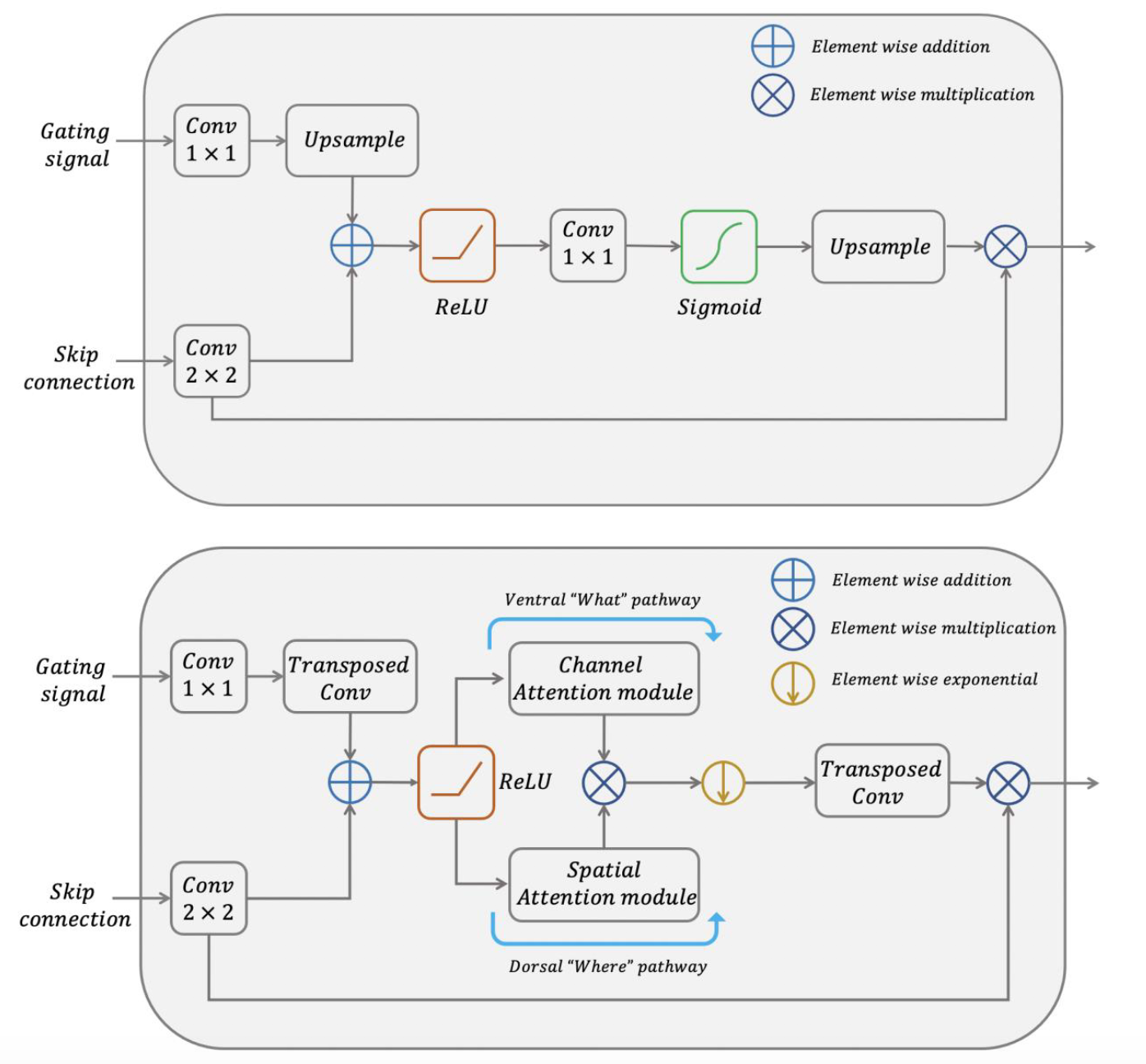}
    \caption{Top: schematic of the additive AG. The gating signal and skip connection are first resized and then combined to form attention coefficients. Multiplication of the original skip connection with the attention coefficients provides spatial context highlighting ROIs. 
Bottom: schematic of the Focus Gate. The gating signal and skip connection are first resized and then combined prior to spatial and object-related feature extraction. The attention coefficients pass through an additional focal filter controlling the degree of background suppression. Finally, multiplying the original skip connection with the attention coefficients provides both spatial and feature context highlighting regions and features of interest.}
    \label{fig:figure_2}
\end{figure}

The structure of the additive AG is illustrated in Fig.~\ref{fig:figure_2} \cite{schlemper2019attention}. This AG receives two inputs, the gating signal and associated skip connection generated at that level. The gating signal originates from the deepest layer of the neural network, where feature representation is the greatest at the cost of significant down-sampling. In contrast, skip connections arise in more superficial layers, where feature representation is coarser, but image resolution is relatively spared. The AG uses contextual information from the gating signal to prune the skip connection, highlighting ROIs and therefore reducing false positive predictions. To accomplish this, the initial stage involves simultaneous upsampling of the gating signal and downsampling of the skip connection to produce equivalent image dimensions enabling element-wise addition. Although computationally more expensive, additive attention has been shown to achieve higher accuracy than multiplicative attention \cite{luong2015effective}. 

The resulting matrix is passed through a ReLU activation, followed by global average pooling along the channel axis and final sigmoid activation, generating a matrix of attention weights, also known as the attention coefficients $i \in [0,1]$. 

The final step is an element-wise multiplication of the upsampled attention coefficients with the original skip connection input, providing spatial context to the skip connection prior to fusion with outputs from the decoder network.

Before describing the FG, illustrated in the bottom of Fig.~\ref{fig:figure_2}, we first describe two of its main components, namely the channel attention module and the spatial attention module.

\subsection{Channel attention module}

The global average pooling operation in the additive AG extracts the spatial context to localise the ROIs. However, by pooling across the channel axis, information conveyed by the channels relating to objects features, such as edges and colour, is lost. On the contrary, by assigning weights along the channel axis, channel interdependencies may be explicitly modelled, enabling networks to better recalibrate the features used for segmentation \cite{hu2018squeeze, fu2019dual, woo2018cbam, gao2019global}. Squeeze-and-excitation (SE) blocks achieve this by initial feature aggregation using global average pooling along the spatial axis, known as the ‘squeeze’ operation, followed by two fully connected layers with ReLU and sigmoid activations producing the ‘excitation’ operation \cite{hu2018squeeze, rundo2019use}. The two fully connected layers involve dimensionality reduction to control model complexity, with implications for computation and performance. Efficient Channel Attention (ECA) \cite{wang2020eca} avoids dimensionality reduction by modelling cross-channel interaction with an adaptive kernel size $k$, defined by:

\begin{equation}
k=\left|\frac{\log _{2}(C)}{\gamma}+\frac{b}{\gamma}\right|_{\text {odd }}
\end{equation}

where $C$ is the channel dimension, while $b$ and $\gamma$ are set to $2$ and $1$, respectively. 

A separate insight incorporated into the Convolutional Block Attention Module (CBAM) for channel attention involves using a global max pooling operation in addition to global average pooling, providing two complementary spatial contexts prior to feature recalibration \cite{woo2018cbam}.

\begin{figure}[ht!]
    \centering
    \includegraphics[width=0.8\textwidth]{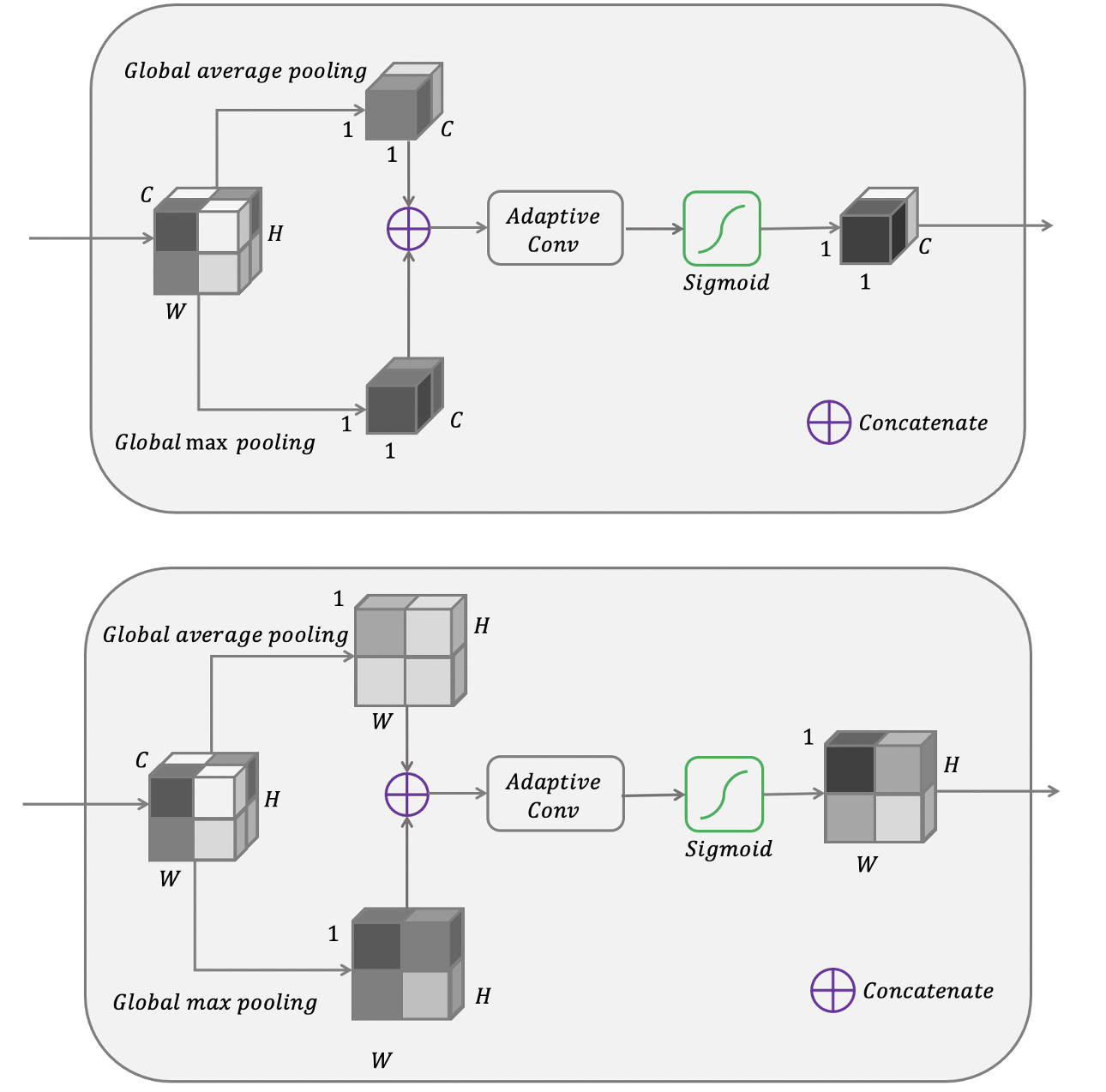}
    \caption{Top: schematic of the channel attention module used in the Focus Gate. Global max pooling and global average pooling generate two complementary spatial contexts prior to channel weighting.
    Bottom: schematic of the spatial attention module used in the Focus Gate. Global max pooling and global average pooling generate two complementary channels contexts prior to spatial context generation.}
    \label{fig:figure_3}
\end{figure}

The channel module used in the FG is illustrated in Fig.~\ref{fig:figure_3}. 

We extend the ideas provided by ECA and CBAM by using initial global average and global max pooling to generate two separate spatial contexts, followed by feature recalibration using an adaptive convolutional kernel size avoiding dimensionality reduction. Finally, a sigmoid activation redistributes the values between $[0, 1]$, generating attention coefficients along the channel axis.

\subsection{Spatial attention module}

Complementary to the channel attention module, spatial attention modules involve feature aggregation along the channel axis \cite{fu2019dual, woo2018cbam, roy2018concurrent}. While dimensional reduction is not an issue for spatial attention modules, the replacement of fully connected layers with a convolutional layer requires an additional kernel size parameter. Larger kernel sizes provide a larger receptive field, with better performance but at the cost of computational efficiency \cite{roy2018concurrent}. The spatial attention module used in the FG is illustrated in Fig.~\ref{fig:figure_3}. 
Again, we extend the ideas provided by ECA and CBAM by using initial global average and global max pooling along the channel axis, generating two separate channel contexts, followed by spatial recalibration with an adaptive convolutional kernel size. In contrast to ECA, the spatial dimension is inversely proportional to the channel dimension, and therefore we modify the original equation and determine kernel size $k$ for the spatial attention module by:

\begin{equation}
k=\left|\frac{\log _{2}\left(C_{\max }+C_{0}-C\right)}{\gamma}+\frac{b}{\gamma}\right|_{\mathrm{odd}}
\end{equation}

where Cmax is the maximum channel dimension of the network, $C_{0}$ is the channel dimension for the first layer, and C is the channel dimension for current layer. The parameters b and $gamma$ are set to 2 and 1, respectively. This provides an efficient compromise by scaling the kernel size in proportion to the input dimension, with larger kernel sizes reserved for larger inputs.

\subsection{Focus gate}

Having introduced both spatial and channel attention modules, in this section we describe the structure of the FG (Fig.~\ref{fig:figure_2}). 

Similar to the attention gate, the gating signal is generated from the deepest layer of the network. The upsampling operation is replaced with a learnable kernel weight using a transposed convolution, but otherwise the skip connection and gating signal are resampled to matching dimensions. Following element-wise addition and non-linear activation, spatial and channel attention coefficients are processed in parallel, analogous to processing of the dorsal ``where'' and ventral ``what'' pathways, respectively, of the two-streams hypothesis for visual processing \cite{goodale1992separate}. The spatial and channel attention coefficients are combined with element-wise multiplication, and passed through a tunable filter involving element-wise exponential parameterised by the focal parameter prior to resampling. 

\begin{figure}[ht!]
    \centering
    \includegraphics[width=0.8\textwidth]{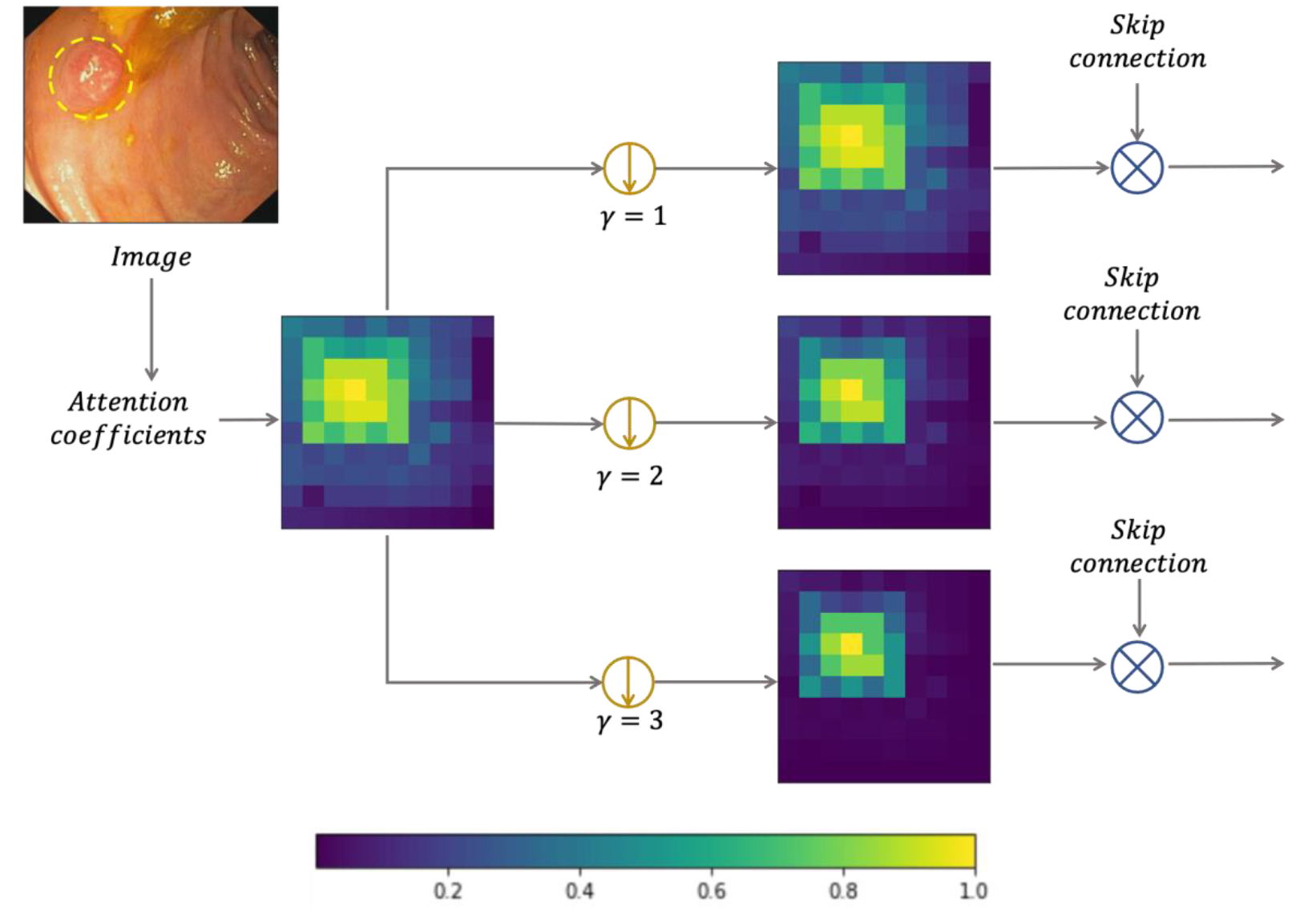}
    \caption{The effect of modifying the focal parameter in the Focus Gate. The brighter colours in the heatmaps are associated with higher activations, here corresponding to the polyp location. Higher focal parameter values lead to increased background suppression. }
    \label{fig:figure_4}
\end{figure}

The concept of a focal parameter originates from work on loss function optimisation, where the contributions of easy examples are downweighed enabling the learning of harder examples \cite{lin2017focal, abraham2019novel}. Here, we apply the focal parameter to the matrix of attention coefficients, enhancing the contrast between foreground and background objects by controlling the degree of background suppression. Following sigmoid activation, all attention coefficient values are redistributed $i \in [0,1]$. This enables higher values of the focal parameter to significantly reduce the weights of irrelevant regions and features, while salient regions and important features are relatively spared. The effect of altering the focal parameter is illustrated in Fig.~\ref{fig:figure_4}. 

Careful tuning of the focal parameter is required, to suppress background regions while preserving attention for borders between foreground and background where attention coefficients take middle values. 

\subsection{Deep supervision}

The vanishing and exploding gradients problems are well-recognised issues with training deep CNNs \cite{glorot2010understanding, pascanu2013difficulty}. The Focus U-Net incorporates two separate, complementary mechanisms to overcome this. Firstly, short-range skip connections, in addition to the long-range skip connections characteristic of the U-Net, allow the error signal to propagate to earlier layers more directly. However, this comes at the cost of computational efficiency, and therefore for the Focus U-Net, we set a compromise by restricting short-range skip connections connections to the input and final output of each layer, providing an efficient implementation while maintaining performance gains. 

In contrast, deep supervision encourages semantic discrimination of intermediate feature maps at each level by assigning a loss to outputs at multiple layers \cite{lee2015deeply, dou20173d}. Equal weighting of outputs produces sub-optimal results due to converging to solutions favouring improved performance of deeper layers at the cost of performance of the final layer. To overcome this, more complicated solutions have been developed, such as multi-scale training \cite{abraham2019novel}, or fine-tuning using a fully connected layer \cite{schlemper2019attention}. To preserve efficiency, here we assign weights w to different output layers according to the equation:

\begin{equation}
w=2^{-(\text {stride length} \times \text {stride width})}
\end{equation}

where the stride length and width refer to the final transposed convolution stride dimensions required to resample the feature map to the original image dimension.

Intuitively, higher weights are therefore assigned to the layers requiring a smaller degree of upsampling, with the greatest weight assigned to the final output, followed by an exponential decrease in weighting with increasing depth of the network. 

\subsection{Hybrid Focal loss}

The training of neural networks is based on solving the optimisation problem defined by the loss function. For semantic segmentation tasks, a popular choice of loss function is the sum of the Dice similarity coefficient (DSC) loss and cross entropy (CE) loss:

\begin{equation}
\mathcal{L}_{\text {total }}=\mathrm{DSC}+\mathcal{L}_{\mathrm{CE}}
\end{equation}

where:

\begin{equation}
\mathrm{DSC}=\frac{2 \mathrm{TP}}{2 \mathrm{TP}+\mathrm{FP}+\mathrm{FN}}
\end{equation}

\begin{equation}
\mathcal{L}_{\mathrm{CE}}(\boldsymbol{y}, \widehat{\boldsymbol{y}})=-(\boldsymbol{y} \log (\widehat{\boldsymbol{y}})+(1-\boldsymbol{y}) \log (1-\widehat{\boldsymbol{y}}))
\end{equation}
where TP, FP and FN refer to true positives, false positives and false negatives respectively, and $\boldsymbol{y}, \boldsymbol{\hat{y}} \in \{0,1\} ^{N}$ where $\boldsymbol{\hat{y}}$ refers to the predicted value and $\boldsymbol{y}$ refers to the ground truth label.

However, with class imbalanced tasks such as polyp segmentation, the resulting segmentation using the Dice loss often leads to high precision but low recall rate \cite{salehi2017tversky}. By weighting false negative predictions more heavily, the Tversky loss improves recall-precision balance:

\begin{equation}
\mathcal{L}_{\text {Tversky }}=1-\mathrm{TI},
\end{equation}
where the Tversky index (TI) is defined as:

\begin{equation}
\mathrm{TI}=\frac{\sum_{I=1}^{N} p_{0 i} g_{0 i}}{\sum_{i=1}^{N} p_{0 i} g_{0 i}+\alpha \sum_{i=1}^{N} p_{0 i} g_{1 i}+\beta \sum_{i=1}^{N} p_{1 i} g_{0 i}}
\end{equation}
 
$p_{0 i}$ is the probability of pixel $i$ belonging to the foreground class and $p_{1 i}$ is the probability of pixel belonging to background class; $g_{0 i}$ is $1$ for foreground and $0$ for background and conversely $g\textsubscript{1i}$ takes values of $1$ for background and $0$ for foreground.

Complementary to weighting the positive and negative examples, applying focal parameters to both the Tversky and cross entropy loss enables the downweighting of background objects in favour of foreground object segmentation, and produces the Focal Tversky loss and Focal loss, respectively \cite{lin2017focal, abraham2019novel}:

\begin{equation}
\mathcal{L}_\text{FTL}=(1-\text{TI})^{\gamma},
\end{equation}

\begin{equation}
\mathcal{L}_\text{FL}=-\alpha\left(1-p_{t}\right)^{y} \cdot \mathcal{L}_\text{CE},
\end{equation}
where $\alpha$ controls the class weights.  

Finally, we define the Hybrid Focal loss (HFL) as the sum of the Focal Tversky loss and Focal loss:

\begin{equation}
\mathcal{L}_\text{HFL}=\mathcal{L}_\text{FTL}+\mathcal{L}_\text{FL}.
\end{equation}

To mitigate suppression of the loss near convergence, we supervise the last layer without the focal parameters \cite{abraham2019novel}. 

\section{Materials and evaluation methods}
\label{sec:MaterialsAndEvalMethods}

\subsection{Dataset descriptions}
\label{sec:Datasets}

We assess the ability of the Focus U-Net to accurately segment polyps using five public datasets containing images of polyps taken during optical colonoscopy: CVC-ClinicDB \cite{bernal2015wm}, Kvasir-SEG \cite{jha2020kvasir}, CVC-ColonDB \cite{tajbakhsh2015automated}, ETIS-Larib PolypDB \cite{silva2014toward} and EndoScene test set (CVC-T) \cite{vazquez2017benchmark}. CVC-ClinicDB, CVC-ColonDB, ETIS-Larib PolypDB and CVC-T were created by the Hospital Clinic, Universidad de Barcelona, Spain, while the Kvasir-SEG database was produced by the Vestre Viken Health Trust, Norway.

The CVC-ClinicDB database consists of 612 frames containing polyps with image resolution 288×368 pixels, generated from 23 video sequences from 13 different patients using standard optical colonoscopy interventions. The Kvasir-SEG database consists of 1000 polyp images collected and verified by experienced gastroenterologists. Images vary in size from 332×487 to 1920×1072 pixels. CVC-ColonDB consists of 300 images of resolution 500×574 pixels obtained from 15 video sequences with a random sample of 20 frames per sequence. ETIS-Larib PolypDB similarly consists of 300 images, with image resolution 1225×966 pixels.  Lastly, CVC-T consists of 182 frames containing polyps from 8 patients derived from either the CVC-ClinicDB and CVC-ColonDB datasets, with image resolutions of 288×384 or 500×574 pixels.

\subsection{Experimental setup and implementation details}
\label{sec:ExpSetup}

For our experiments, we use the Medical Imaging Segmentation with Convolutional Neural Networks (MIScnn) open-source Python library \cite{muller2021miscnn}. For all datasets, images and associated ground truth masks are provided in the png file format. For the Kvasir-SEG dataset, we resize all images to 512×512 pixels following pre-processing methods used in previous models \cite{fan2020pranet, huang2021hardnet}, but otherwise resize images to 288×384 pixels for all other datasets. Pixel values are normalised to [0, 1] using the z-score. We perform full-image analysis with a batch size of 16, except for the Kvasir-SEG dataset where the large image sizes required the batch size to be reduced to 8. We use the Focus U-Net architecture as described previously with a final softmax activation layer.

For the ablation studies, we use the CVC-ClinicDB dataset, with five-fold cross validation using random assignment. We evaluate the baseline performance of the U-Net \cite{ronneberger2015u} and Attention U-Net \cite{schlemper2019attention}, and sequentially assess the performance with subsequent additions of the Focus Gate, Hybrid Focal loss, short-range skip connections and deep supervision. We select values for the focal parameter $\gamma \in [1, 3]$. Model parameters are initialised with Xavier initialisation, and each model is trained for 100 epochs using Stochastic Gradient Descent with Nesterov momentum ($\mu=0.99$). We set the initial learning rate at 0.01, and follow a polynomial learning decay rate schedule \cite{isensee2021nnu}:

\begin{equation}
\left(\frac{1-\text { epoch }}{\text { epoch }_{\max }}\right)^{0.9}
\end{equation}

For fairer comparison, we do not apply any data augmentation techniques at this stage.

In contrast, when attempting for state-of-the-art results on the CVC-ClinicDB dataset, we train our final model using five-fold cross validation for 500 epochs and use the following data augmentation techniques: scaling, rotation, elastic deformation, mirror and gamma transformations. For all other datasets, we follow the single train-test split used in \cite{fan2020pranet, huang2021hardnet} for evaluation on the Kvasir-SEG dataset, as well as for evaluation on the combination of all five public datasets, and train each model for 1000 epochs with the same data augmentation settings. 

For the Hybrid Focal loss, we follow the optimal hyperparameters reported in the original studies. We set $\alpha=0.3$ and $\beta=0.7$ for the Tversky index, $\alpha = 0.25$, $\gamma = 2$ for the Focal loss and $\alpha=0.3$, $\beta=0.7$ and $\gamma = 3/4$ for the Focal Tversky loss \cite{lin2017focal, abraham2019novel, salehi2017tversky}

For all cases, the validation loss is evaluated at the end of each epoch, and the model with the lowest validation loss is selected as the final model. All experiments are programmed using Keras with TensorFlow backend and trained with NVIDIA P100 GPUs, with CUDA version 10.2 and cuDNN version 7.6.5. Source code is available at: \url{https://github.com/mlyg/Focus-U-Net}

\subsection{Evaluation metrics}
\label{sec:EvalMetrics}

To assess segmentation accuracy, we follow recommendations from \cite{jha2020kvasir}, and use DSC and intersection over union (IoU) as the two main metrics. DSC is previously defined in Eq. (5), and IoU is defined as:

\begin{equation}
\mathrm{IoU}=\frac{\mathrm{TP}}{\mathrm{TP}+\mathrm{FP}+\mathrm{FN}}
\end{equation}

The IoU metric penalises single instances of poor pixel classification more heavily than DSC, providing similar but complementary perspectives on assessing segmentation accuracy. We further assess recall and precision:

\begin{equation}
\text { Recall }=\frac{\mathrm{TP}}{\mathrm{TP}+\mathrm{FN}}
\end{equation}

\begin{equation}
\text { Precision }=\frac{\mathrm{TP}}{\mathrm{TP}+\mathrm{FP}}
\end{equation}

In the context of polyp segmentation, recall, also known as sensitivity, measures the proportion of the pixels corresponding to the polyp that are correctly identified. In contrast, precision, also known as the positive predictive value, measures the proportion of pixels correctly labelled as representing the polyp. While both are accounted for in the DSC metric, measuring recall and precision provides additional information regarding the false positive and false negative rates.

\section{Experimental results}
\label{sec:Results}

We first perform a series of ablation studies to evaluate individual components of the Focus U-Net, followed by separate evaluations on the CVC-ClinicDB and Kvasir-SEG datasets, and finally evaluate against a test set combining five public polyp datasets.

The results from the ablation study are shown in Table~\ref{tab:table1}. 
 
\begin{table}[]
\centering
\caption{Results from training on the CVC-ClinicDB for the U-Net, Attention U-Net and Focus U-Net. The best result is seen with the addition of the Focus Gate, short skip connections, deep superivsion and use of the Hybrid Focal loss.}
\hspace*{-1.8cm}
\scalebox{0.55}{
\begin{tabular}{lcccccccc}
\multicolumn{1}{c}{Model} & Loss function & Focal parameter $\gamma$ & Short skip connections & Deep supervision & mDSC       & mIoU       & Recall      & Precision  \\
\hline
U-Net                     & DSC + CE      & -               & {\color[HTML]{333333} \xmark}                      & {\color[HTML]{333333} \xmark}                & 0.828±0.021 & 0.747±0.022 & 0.817±0.024 & 0.877±0.021 \\
Attention U-Net            & DSC + CE      & -               & {\color[HTML]{333333} \xmark}                      & {\color[HTML]{333333} \xmark}                & 0.801±0.019 & 0.705±0.023 & 0.799±0.012  & 0.844±0.030 \\
Focus U-Net               & DSC + CE      & 1               & {\color[HTML]{333333} \xmark}                      & {\color[HTML]{333333} \xmark}                & 0.838±0.018 & 0.755±0.018 & 0.833±0.016  & 0.876±0.028 \\
Focus U-Net               & DSC + CE      & 1.25            & {\color[HTML]{333333} \xmark}                      & {\color[HTML]{333333} \xmark}                & 0.844±0.011 & 0.762±0.011 & 0.844±0.018  & 0.876±0.020 \\
Focus U-Net               & DSC + CE      & 1.5             & {\color[HTML]{333333} \xmark}                      & {\color[HTML]{333333} \xmark}                & 0.842±0.025 & 0.755±0.027 & 0.845±0.038  & 0.866±0.016 \\
Focus U-Net               & DSC + CE      & 2               & {\color[HTML]{333333} \xmark}                      & {\color[HTML]{333333} \xmark}                & 0.817±0.025 & 0.728±0.030 & 0.817±0.023  & 0.859±0.026 \\
Focus U-Net               & DSC + CE      & 3               & {\color[HTML]{333333} \xmark}                      & {\color[HTML]{333333} \xmark}                & 0.825±0.022 & 0.736±0.022 & 0.820±0.035  & 0.863±0.022 \\
Focus U-Net               & DSC + CE      & 1.25            & {\color[HTML]{333333} \cmark}                      & {\color[HTML]{333333} \xmark}                & 0.867±0.018 & 0.800±0.011 & 0.852±0.023  & \textbf{0.908±0.017} \\
Focus U-Net               & HFL           & 1.25            & {\color[HTML]{333333} \cmark}                      & {\color[HTML]{333333} \xmark}                & 0.869±0.013 & 0.797±0.014 & 0.870±0.017  & 0.892±0.010 \\
Focus U-Net               & HFL           & 1.25            & {\color[HTML]{333333} \cmark}                      & {\color[HTML]{333333} \cmark}                & \textbf{0.875±0.016} & \textbf{0.801±0.017} & \textbf{0.878±0.013}  & 0.889±0.018
\end{tabular}}
\label{tab:table1}
\end{table}

Performance gains are observed with successive addition of each component, and with all components present there is a significant improvement with a DSC score of $0.875 \pm 0.016$ compared to the U-Net ($0.828 \pm 0.021$) and Attention U-Net ($0.801 \pm 0.019$). 

The results for the CVC-ClinicDB dataset are shown in Table~\ref{tab:table2}. 

\begin{table}[]
\centering
\caption{Results for the CVC-ClinicDB dataset. Boldface numbers denote the highest values for each metrics.}
\scalebox{0.9}{
\begin{tabular}{lllll}
Model                            & mDSC           & mIoU           & Recall         & Precision      \\
\hline
FCN-8S \cite{akbari2018polyp}                           & 0.810          & -              & 0.748          & 0.883          \\
Multi-scale patch-based CNN \cite{banik2020multi}      & 0.813          & -              & 0.786          & 0.809          \\
FCN \cite{li2017colorectal}                              & 0.830          & -              & 0.773          & 0.900          \\
MultiResUNet \cite{ibtehaz2020multiresunet}                     & -              & 0.821          & -              & -              \\
cGAN \cite{poorneshwaran2019polyp}                             & 0.872          & 0.795          & -              & -              \\
U-Net \cite{ronneberger2015u}                            & 0.878          & 0.788          & 0.787          & 0.933          \\
Multiple encoder-decoder network \cite{nguyen2018colorectal} & 0.889          & 0.894          & -              & -              \\
PraNet \cite{fan2020pranet}                           & 0.898          & 0.840          & -              & -              \\
PolypSegNet \cite{mahmud2021polypsegnet}                     & 0.915          & 0.862          & 0.911          & \textbf{0.962} \\
ResUNet++ with CRF \cite{jha2021comprehensive}              & 0.920          & 0.890          & 0.939          & 0.846          \\
Double U-Net \cite{jha2020doubleu}                     & 0.924          & 0.861          & 0.846          & 0.959          \\
Focus U-Net                      & \textbf{0.941} & \textbf{0.893} & \textbf{0.956} & 0.930         
\end{tabular}}
\label{tab:table2}
\end{table}

The Focus U-Net achieves state-of-the-art results with a mDSC score of $0.941$ and a mIoU score of $0.893$, outperforming the ResUNet++ with Conditional Random Field (CRF) and DoubleU-Net. Focus U-Net also has the best recall-precision balance, while PolypSegNet achieves the highest precision at the cost of recall, and conversely the ResUNet++ with CRF achieves high recall at the cost of precision.

Next, we evaluate our model on the Kvasir-SEG dataset. The results are shown in Table~\ref{tab:table3}.
The Focus U-Net achieves state-of-the-art results with a mDSC score of 0.910 and mIoU score of 0.845. The highest mIoU is achieved by HarDNet-MSEG \cite{huang2021hardnet}.

\begin{table}[]
\centering
\caption{Results for the Kvasir-SEG dataset. Boldface numbers denote the highest values for each metrics}
\scalebox{0.9}{
\begin{tabular}{lllll}
Model                           & mDSC           & mIoU           & Recall         & Precision      \\
\hline
U-Net \cite{ronneberger2015u}                           & 0.715          & 0.433          & 0.631          & 0.922          \\
Double U-Net \cite{jha2020doubleu}                    & 0.813          & 0.733          & 0.840          & 0.861          \\
FCN8 (VGG16 backbone) \cite{long2015fully}          & 0.831          & 0.737          & 0.835          & 0.882          \\
PSPNet (ResNet50 backbone) \cite{zhao2017pyramid}     & 0.841          & 0.744          & 0.836          & 0.890          \\
HRNet \cite{wang2020deep}                          & 0.845          & 0.759          & 0.859          & 0.878          \\
ResUNet++ with CRF \cite{jha2021comprehensive}              & 0.851          & 0.833          & 0.876          & 0.823          \\
DeepLabv3+ (ResNet101 backbone) \cite{chen2018encoder} & 0.864          & 0.786          & 0.859          & 0.906          \\
U-Net (ResNet34 backbone) \cite{ronneberger2015u}       & 0.876          & 0.810          & 0.944          & 0.862          \\
PolypSegNet \cite{mahmud2021polypsegnet}                   & 0.887          & 0.826          & \textbf{0.925} & \textbf{0.917} \\
HarDNet-MSEG \cite{huang2021hardnet}                    & 0.904          & \textbf{0.848} & 0.923          & 0.907          \\
Focus U-Net                     & \textbf{0.910} & 0.845          & 0.916          & \textbf{0.917}
\end{tabular}}
\label{tab:table3}
\end{table}

Finally, Table~\ref{tab:table4} shows the results for the evaluation on five public polyp datasets. 

\begin{table}[]
\centering
\caption{Results for evaluation on five public polyp datasets. The combined score takes into account the number of images used for each dataset, enabling cross dataset comparisons.}
\scalebox{0.55}{
\begin{tabular}{lcccccccccccc}
\hline
\multicolumn{13}{c}{Dataset (number of images)}                                                                                                                                                                                                                                                                                                                                                                                                                                                                                                                 \\ \hline
\multicolumn{1}{l|}{}             & \multicolumn{2}{c|}{\begin{tabular}[c]{@{}c@{}}CVC-ClinicDB\\    (61)\end{tabular}} & \multicolumn{2}{c|}{\begin{tabular}[c]{@{}c@{}}CVC-ColonDB\\    (380)\end{tabular}} & \multicolumn{2}{c|}{\begin{tabular}[c]{@{}c@{}}ETIS-LaribPolypDB\\    (196)\end{tabular}} & \multicolumn{2}{c|}{\begin{tabular}[c]{@{}c@{}}CVC-T\\    (60)\end{tabular}} & \multicolumn{2}{c|}{\begin{tabular}[c]{@{}c@{}}Kvasir-SEG\\    (100)\end{tabular}} & \multicolumn{2}{c}{\begin{tabular}[c]{@{}c@{}}Combined\\    Score\end{tabular}} \\ \hline
\multicolumn{1}{l|}{Model}        & mDSC                           & \multicolumn{1}{c|}{mIoU}                          & mDSC                           & \multicolumn{1}{c|}{mIoU}                          & mDSC                              & \multicolumn{1}{c|}{mIoU}                             & mDSC                       & \multicolumn{1}{c|}{mIoU}                       & mDSC                          & \multicolumn{1}{c|}{mIoU}                          & mDSC                                   & mIoU                                   \\
\multicolumn{1}{l|}{SFA \cite{fang2019selective}}          & 0.700                          & \multicolumn{1}{c|}{0.607}                         & 0.469                          & \multicolumn{1}{c|}{0.347}                         & 0.297                             & \multicolumn{1}{c|}{0.217}                            & 0.467                      & \multicolumn{1}{c|}{0.329}                      & 0.723                         & \multicolumn{1}{c|}{0.611}                         & 0.476                                  & 0.367                                  \\
\multicolumn{1}{l|}{U-Net++ \cite{zhou2019unet++}}      & 0.794                          & \multicolumn{1}{c|}{0.729}                         & 0.483                          & \multicolumn{1}{c|}{0.410}                         & 0.401                             & \multicolumn{1}{c|}{0.344}                            & 0.707                      & \multicolumn{1}{c|}{0.624}                      & 0.821                         & \multicolumn{1}{c|}{0.743}                         & 0.546                                  & 0.476                                  \\
\multicolumn{1}{l|}{ResUNet++ \cite{jha2019resunet++}}    & 0.796                          & \multicolumn{1}{c|}{0.796}                         & -                              & \multicolumn{1}{c|}{-}                             & -                                 & \multicolumn{1}{c|}{-}                                & -                          & \multicolumn{1}{c|}{-}                          & 0.813                         & \multicolumn{1}{c|}{0.793}                         & -                                      & -                                      \\
\multicolumn{1}{l|}{U-Net \cite{ronneberger2015u}}        & 0.823                          & \multicolumn{1}{c|}{0.755}                         & 0.512                          & \multicolumn{1}{c|}{0.444}                         & 0.398                             & \multicolumn{1}{c|}{0.335}                            & 0.710                      & \multicolumn{1}{c|}{0.627}                      & 0.818                         & \multicolumn{1}{c|}{0.746}                         & 0.561                                  & 0.493                                  \\
\multicolumn{1}{l|}{PraNet \cite{fan2020pranet}}       & 0.899                          & \multicolumn{1}{c|}{0.849}                         & 0.709                          & \multicolumn{1}{c|}{0.640}                         & 0.628                             & \multicolumn{1}{c|}{0.567}                            & 0.871                      & \multicolumn{1}{c|}{0.797}                      & 0.898                         & \multicolumn{1}{c|}{0.840}                         & 0.740                                  & 0.675                                  \\
\multicolumn{1}{l|}{HarDNet-MSEG \cite{huang2021hardnet}} & 0.932                          & \multicolumn{1}{c|}{0.882}                         & 0.731                          & \multicolumn{1}{c|}{0.660}                         & 0.677                             & \multicolumn{1}{c|}{0.613}                            & 0.887                      & \multicolumn{1}{c|}{0.821}                      & \textbf{0.912}                & \multicolumn{1}{c|}{\textbf{0.857}}                & 0.768                                  & 0.702                                  \\
\multicolumn{1}{l|}{Focus U-Net}  & \textbf{0.938}                 & \multicolumn{1}{c|}{\textbf{0.889}}                & \textbf{0.878}                 & \multicolumn{1}{c|}{\textbf{0.804}}                & \textbf{0.832}                    & \multicolumn{1}{c|}{\textbf{0.757}}                   & \textbf{0.920}             & \multicolumn{1}{c|}{\textbf{0.860}}             & 0.910                         & \multicolumn{1}{c|}{0.853}                         & \textbf{0.878}                         & \textbf{0.809}                        
\end{tabular}}
\label{tab:table4}
\end{table}

The Focus U-Net achieves the highest score across four of the five datasets with a mDSC of $0.938$ and mIoU of $0.889$ on the CVC-ClinicDB, mDSC of $0.878$ and mIoU of $0.804$ for CVC-ColonDB, mDSC of $0.832$ and mIoU of $0.757$ for ETIS-LaribPolypDB, mDSC of $0.920$ and mIoU of $0.860$ for CVC-T and mDSC of $0.910$ and mIoU of $0.853$ for the Kvasir-SEG dataset. Importantly, the combined score takes into account the relative number of images of each dataset, and the Focus U-Net achieves a mDSC of $0.878$ and mIoU of $0.702$, a $14\%$ increase in mDSC over the previous state-of-the-art (HarDNet-MSEG, mDSC = $0.768$) and $15\%$ increase in mIoU (HarDNet-MSEG, mIoU = $0.702$). The greatest improvements are observed in the most challenging datasets, namely the CVC-ColonDB and ETIS-LaribPolypDB datasets. For CVC-ColonDB, a $20\%$ increase in mDSC is observed over the previous state-of-the-art (HarDNet-MSEG, mDSC = $0.731$) and an $87\%$ increase in mDSC compared to the previously top performing Selective Feature Aggregation (SFA) model. Even more significantly, for ETIS-LaribPolypDB, a $23\%$ increase in mDSC is observed over the previous state-of-the-art (HarDNET-MSEG, mDSC = $0.677$) and a $180\%$ increase in mDSC compared to SFA. 

\begin{figure}[ht!]
    \centering
    \includegraphics[width=\textwidth]{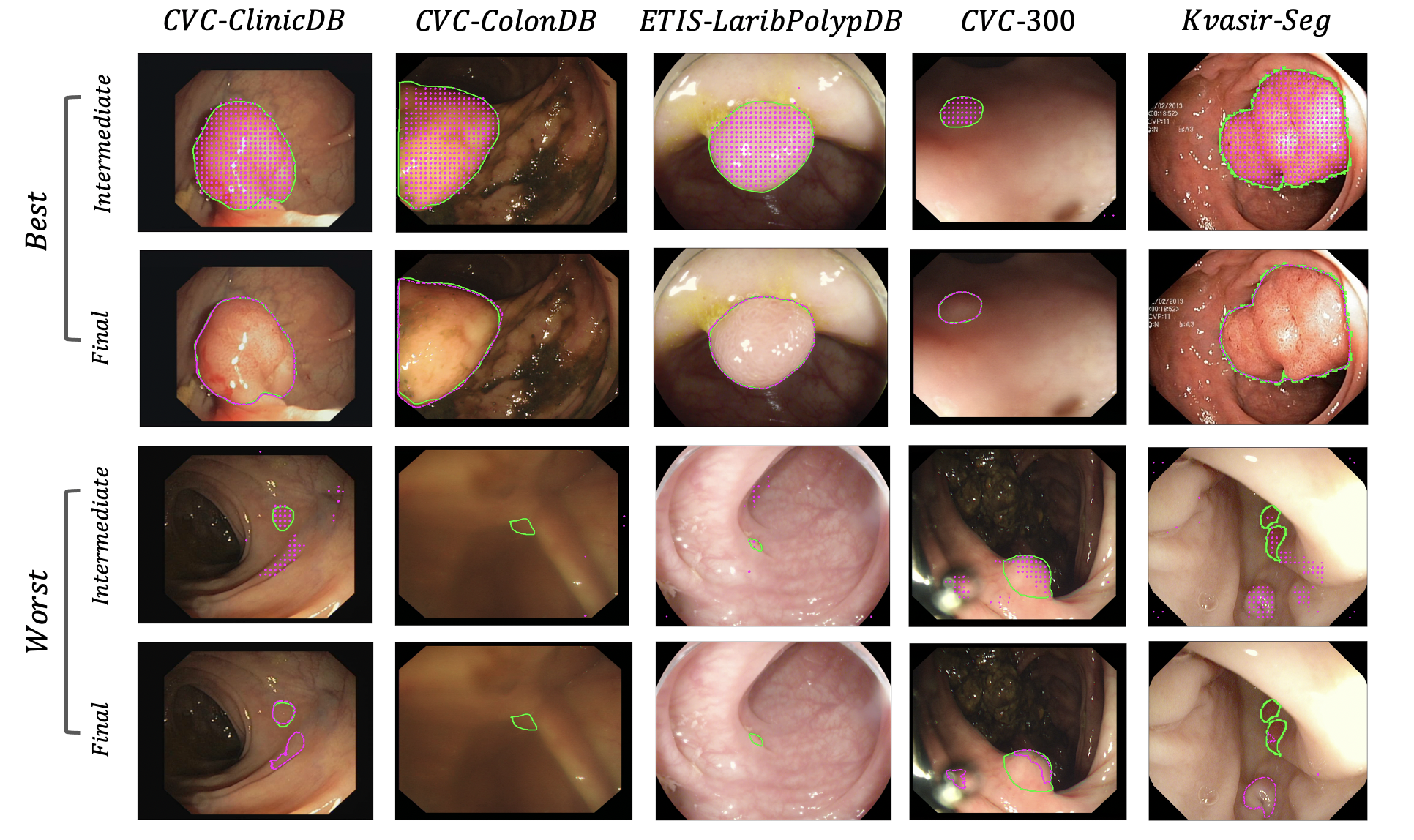}
    \caption{Examples of the best and worst cases in terms of DSC from each of the five public datasets. The solid green line represents the ground truth mask. The dashed magenta line corresponds to the predictions yielded by the Focus U-Net. The intermediate prediction is derived from the deepest layer of the network, while the final prediction is used for evaluation.}
    \label{fig:figure_5}
\end{figure}

Examples of polyp segmentations for the five datasets are shown in Fig.~\ref{fig:figure_5}. 

The accuracy of segmentations obtained from the intermediate layer highlights the ability for the deepest layers to localise the polyp effectively. The Focus U-Net generalises well with consistently accurate segmentations across all datasets. For the images corresponding to the poorest segmentation quality, these are either objectively challenging polyps to identify, or in many cases poor-quality images such as in the CVC-ColonDB example.

\section{Discussion and conclusion}
\label{sec:Discussion}

In this paper, we introduce a novel dual attention-gated U-Net architecture, named the Focus U-Net, which uses a Focus Gate to encourage learning of salient regions combined with a focal parameter controlling suppression of irrelevant background regions. Moreover, with the additions of short-range skip connections and deep supervision, as well as optimisation based on the Hybrid Focal loss, the Focus U-Net outperforms the state-of-the-art results across five public polyp datasets. Importantly, the proposed architecture performs consistently well across all datasets, demonstrating an ability to generalise to unseen data from different datasets. Visualising the resulting polyp segmentations confirms the segmentation quality, with poorer segmentations associated with a combination of either poorer image quality or objectively more challenging polyps to identify.

The proposed Focus U-Net is the latest addition to lightweight, yet accurate, polyp segmentation models, achieving state-of-the-art results with a mDSC of $0.878$ and mIoU score of $0.809$ when evaluated on the combination of five public datasets, a $14\%$ and $15\%$ improvement over the previous state-of-the-art results from HarDNet-MSEG with a mDSC of $0.768$ and mIoU of $0.702$, respectively.

While these results are promising, it is important to determine whether such a model may be applied in clinical practice. Given that colonoscopy involves recordings of live video, a model with a fast inference time is required to process images in real-time. Accordingly, the Focus U-Net architecture is efficiently designed, with both efficient channel and spatial attention mechanisms, as well as a lightweight U-Net backbone \cite{wang2021hybrid}. With polyp miss rates as high as $26\%$ reported for small adenomas \cite{kim2017miss}, the primary advantage of AI-assisted colonoscopy is to aid clinicians in reducing polyp miss-rate detection. However, a secondary advantage with segmentation-based computer-aided detection is providing an accurate and operator-independent estimate of the polyp size; an important factor in guiding biopsy decisions that may be required during colonoscopy. 

There are several limitations associated with our current study. Firstly, the datasets used to train our model consist of images all containing polyps, in contrast to in practice, where the majority of live video data would not contain a polyp. However, in terms of model training, it has been observed that training with images in the absence of polyps results in poorer generalisation \cite{brandao2018towards}. In terms of model performance, we would expect a higher false positive rate. This is not as undesirable as the converse of a high false negative rate, because the purpose of the computer-aided system is to focus the operator to attend to highlighted regions that may contain missed polyps. 

While colonoscopy remains the gold-standard for investigating suspected CRC, CT virtual colonography is a relatively newer method for bowel cancer screening that offers non-invasive visualisation of the colon \cite{summers2005computed}. The flexibility of our model does not restrict usage to polyps in visible light and is equally applicable for polyp detection using CT colonography. In fact, the scope for using the Focus U-Net architecture is not limited for colonic polyps, and is designed specifically for any image segmentation problem where there is the issue of class imbalance and requirement for efficiency.

\section*{Acknowledgements}
This work was partially supported by The Mark Foundation for Cancer Research and Cancer Research UK Cambridge Centre [C9685/A25177], the CRUK National Cancer Imaging Translational Accelerator (NCITA) [C42780/A27066] and the Wellcome Trust Innovator Award [RG98755].
Additional support was also provided by the National Institute of Health Research (NIHR) Cambridge Biomedical Research Centre [BRC-1215-20014] and the Cambridge Mathematics of Information in Healthcare (CMIH) [funded by the EPSRC grant EP/T017961/1].
The views expressed are those of the authors and not necessarily those of the NHS, the NIHR, or the Department of Health and Social Care.

CBS in addition acknowledges support from the Leverhulme Trust project on `Breaking the non-convexity barrier', the Philip Leverhulme Prize, the Royal Society Wolfson Fellowship, the EPSRC grants EP/S026045/1, EP/N014588/1, European Union Horizon 2020 research and innovation programmes under the Marie Skodowska-Curie grant agreement No. 777826 NoMADS and No. 691070 CHiPS, the Cantab Capital Institute for the Mathematics of Information and the Alan Turing Institute.

This work was performed using resources provided by the Cambridge Service for Data Driven Discovery (CSD3) operated by the University of Cambridge Research Computing Service (\url{www.csd3.cam.ac.uk}), provided by Dell EMC and Intel using Tier-2 funding from the Engineering and Physical Sciences Research Council (capital grant EP/P020259/1), and DiRAC funding from the Science and Technology Facilities Council (\url{www.dirac.ac.uk}).

\bibliographystyle{elsarticle-num}


\bibliography{biblio.bib}







\end{document}